\documentclass[12pt]{article}
\usepackage{amssymb}
\usepackage{rotate}
\usepackage{epsfig}
\usepackage{cite}
\usepackage{graphicx}
\begin{document}
\begin{flushright}
\end{flushright}
\vspace*{20mm}
\begin{center}
{\LARGE\bf $Z\gamma\gamma\gamma\to 0$ processes in {\tt SANC}.}
\vspace*{15mm}

{\bf D.~Bardin, L.~Kalinovskaya, E.~Uglov}
\vspace*{10mm}

{\normalsize{\it
Dzhelepov Laboratory for Nuclear Problems, JINR,   \\
ul. Joliot-Curie 6, RU-141980 Dubna, Russia       }}
\vspace*{20mm}
\end{center}
\begin{abstract}
\noindent
In this paper we describe the analytic and numerical evaluation of the $\gamma\gamma\to \gamma Z$ process 
cross section and the $Z\to \gamma\gamma\gamma$ decay rate within the {\tt SANC} system multi-channel approach 
at the one-loop accuracy level with all masses taking into account. The corresponding package 
for numeric calculations is presented. 
For checking of the results correctness we make a comparison with other independent calculations.
\end{abstract}
\vspace*{20mm}
\footnoterule
{\footnotesize
E-mails: bardin@nusun.jinr.ru, kalinov@nusun.jinr.ru, corner@nusun.jinr.ru}
\clearpage
\section{\label{SEC1} Introduction}

 The article describes the implementation into the system {\tt SANC} \cite{1,2,3} the scattering process 
\begin{eqnarray} 
\label{Reaction}
\gamma \gamma \to \gamma Z, 
\end{eqnarray}
(see \cite{4,5,6}) and the decay 
\begin{eqnarray} 
\label{Decay}
Z \to \gamma \gamma \gamma,
\end{eqnarray}
 (see \cite{7,8}) in the Standard Model (SM) at the one-loop level of accuracy in $R_{\xi}$-gauge with taking into account  
of all masses ($Z$ boson and internal ones).
The processes is interesting from an educational point of view, since the calculation of the cross section 
and the width involves only loops and does not contain diagrams of the tree-level and bremsstrahlung, 
and also exploits the multi-channel approach. 
The work is done in the framework of the 4-bosons processes sector extension in the {\tt SANC} 
system~\cite{9,10}.

 In section \ref{SEC2} the multi-channel approach of diagrams calculation is described, 
when all the particles participating in the process are considered as incoming $Z\gamma\gamma\gamma\to 0$. 
All the one-loop diagrams, as well as their corresponding amplitudes in terms of Lorenz expressions 
(based on the constructed basis) and scalar form factors are discussed, as well as the structure of these 
expressions and a proof of zero axial part in the fermionic loop contribution to these processes through 
the application a special sequence of Shouten identities.

 In section \ref{SEC3} we discuss the helicity amplitudes, resulting in a chosen channel of the process
~(\ref{Reaction}) or the decay~(\ref{Decay}) as well as formulas for the cross section and the decay width, respectively.

\begin{figure}[!h]
\begin{center}
\includegraphics[width=9cm,angle=270]{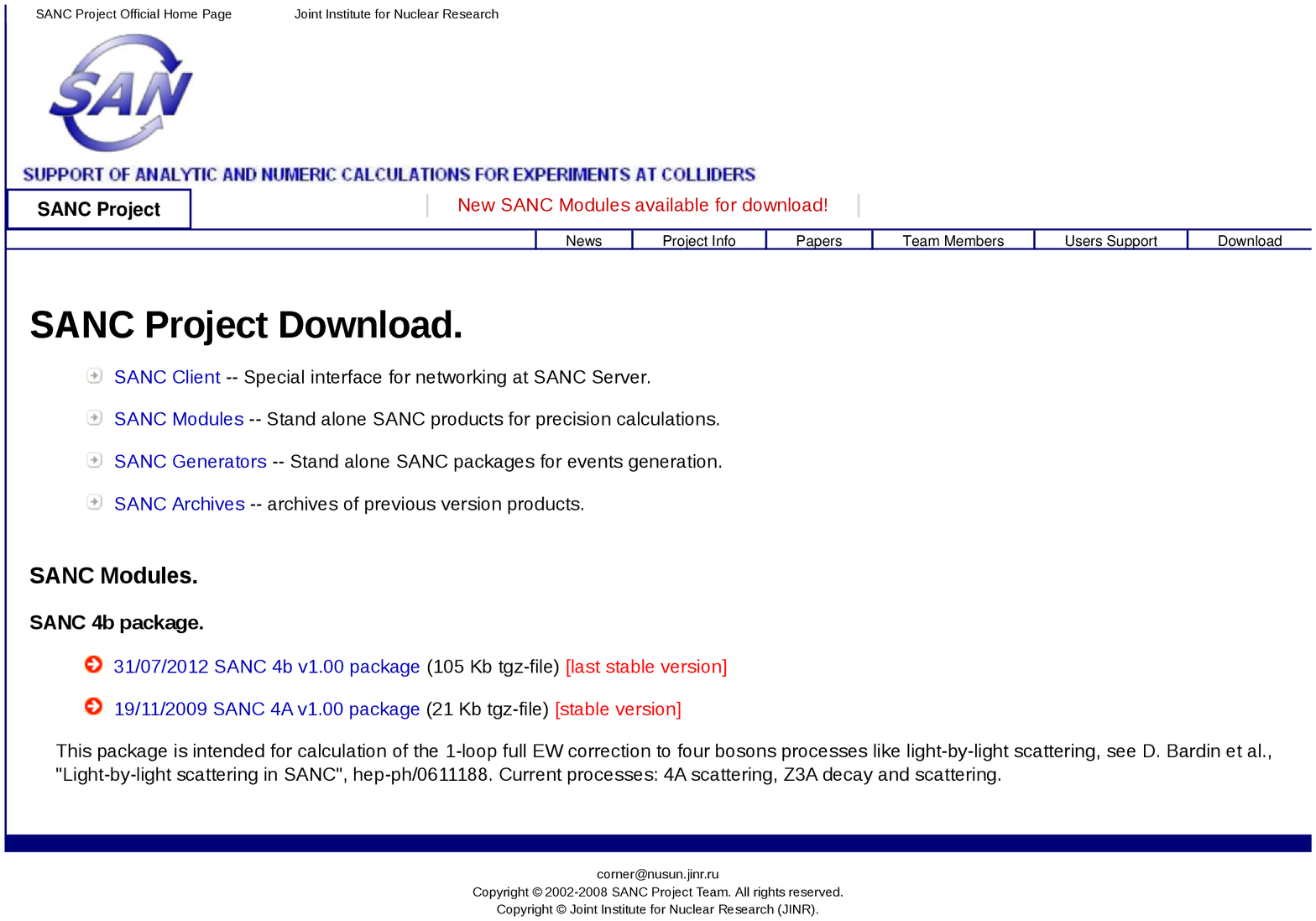}
\end{center}
\caption{{\tt SANC} modules download web site~\cite{3}.\label{FIG1}}
\end{figure}

 In section \ref{SEC4} the package for numerical computations is described. It is created on the basis 
of analytic calculations in the {\tt SANC} environment. There are given the technical instructions and 
the contents of the package, and also the control flags are described.
The sanc\_4b\_v1.00 package contains the processes $\gamma \gamma \to \gamma \gamma$, 
$\gamma \gamma \to \gamma Z$,~(\ref{Reaction}) and $Z \to \gamma \gamma \gamma$,~(\ref{Decay}), see (Fig.\ref{FIG1}),

In section \ref{SEC5} we present the numerical results of the sanc\_4b\_v1.00 package
and comparisons with the known in world literature ones.

In the conclusion we briefly summarize the results of the work.

\section{\label{SEC2} Precomputation level, channel $Z\gamma\gamma\gamma\to 0$}

In the {\tt SANC} system the basic concept of the analytical calculations is precomputation of vacuum
building blocks, namely diagrams, in which all external particles are considered to be incoming 
and not lying on the mass shell. 

These are the building blocks that can be used as the elements in the calculation of real processes 
in relevant channels by means of transformation of external particles momenta and replacing the squares of 
momenta by the squares of masses. 
Consider this concept on the example of the process $Z\gamma\gamma\gamma\to 0$.

The process at the one-loop level of accuracy is described by two blocks of the diagrams with a fermionic
and bosonic propagators, respectively. Their calculation can be made independently.

\begin{figure}[ht!]
\begin{center}
\includegraphics[width=3.3cm]{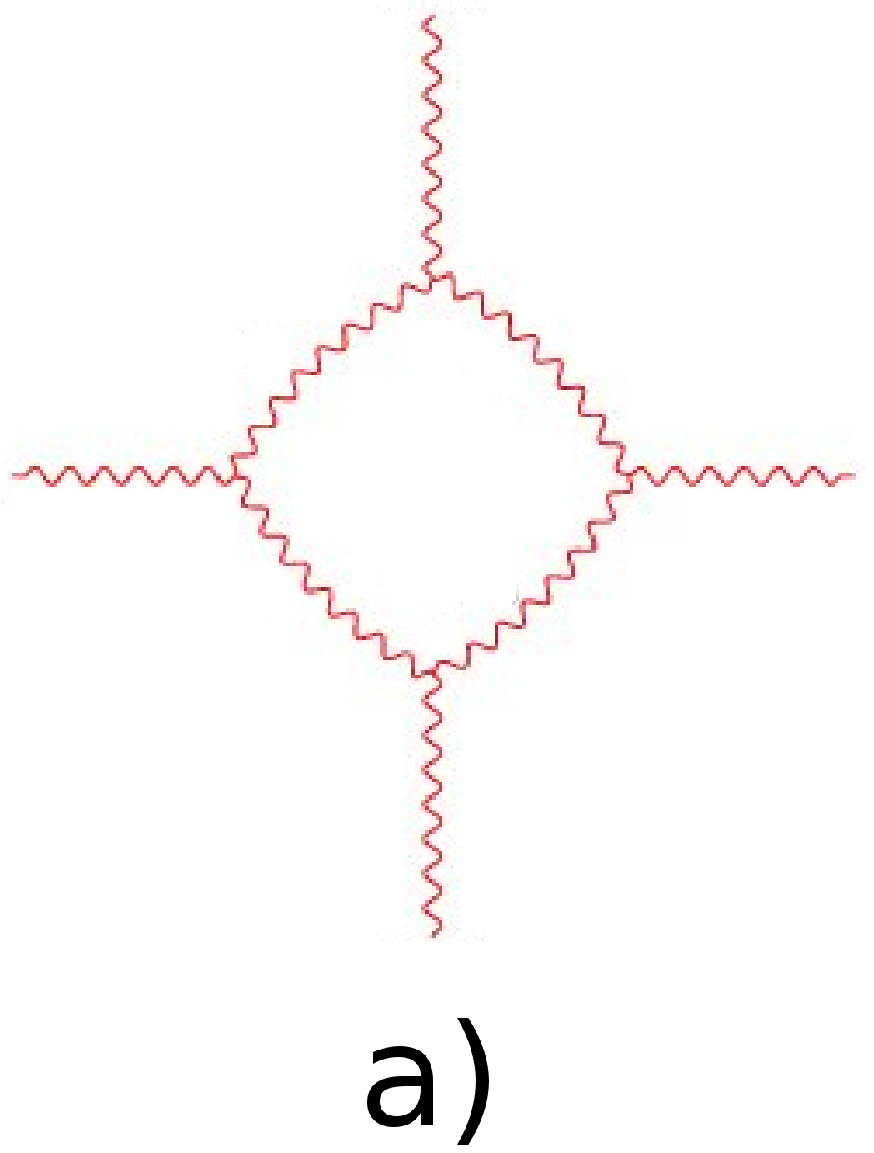}
\includegraphics[width=3.3cm]{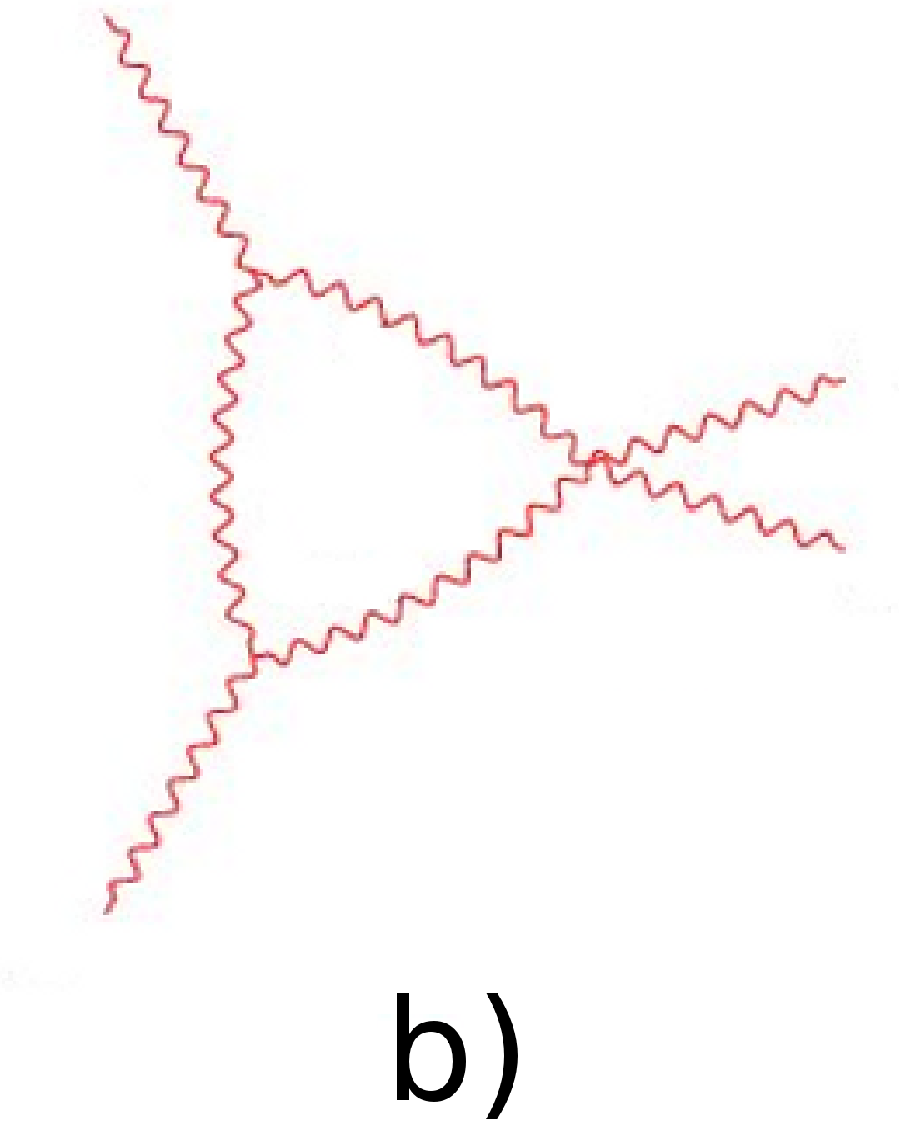}
\includegraphics[width=3.3cm]{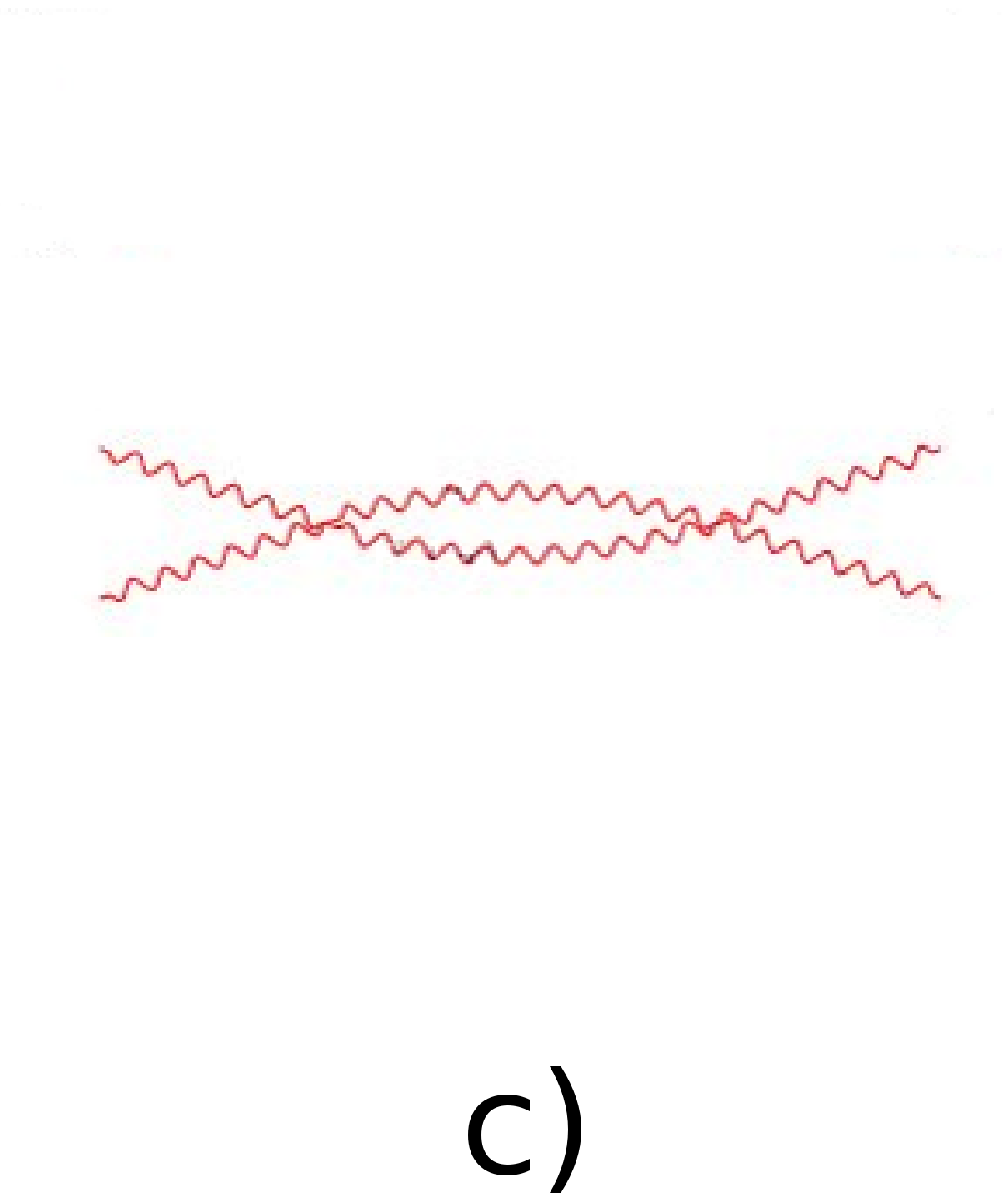}
\includegraphics[width=3.3cm]{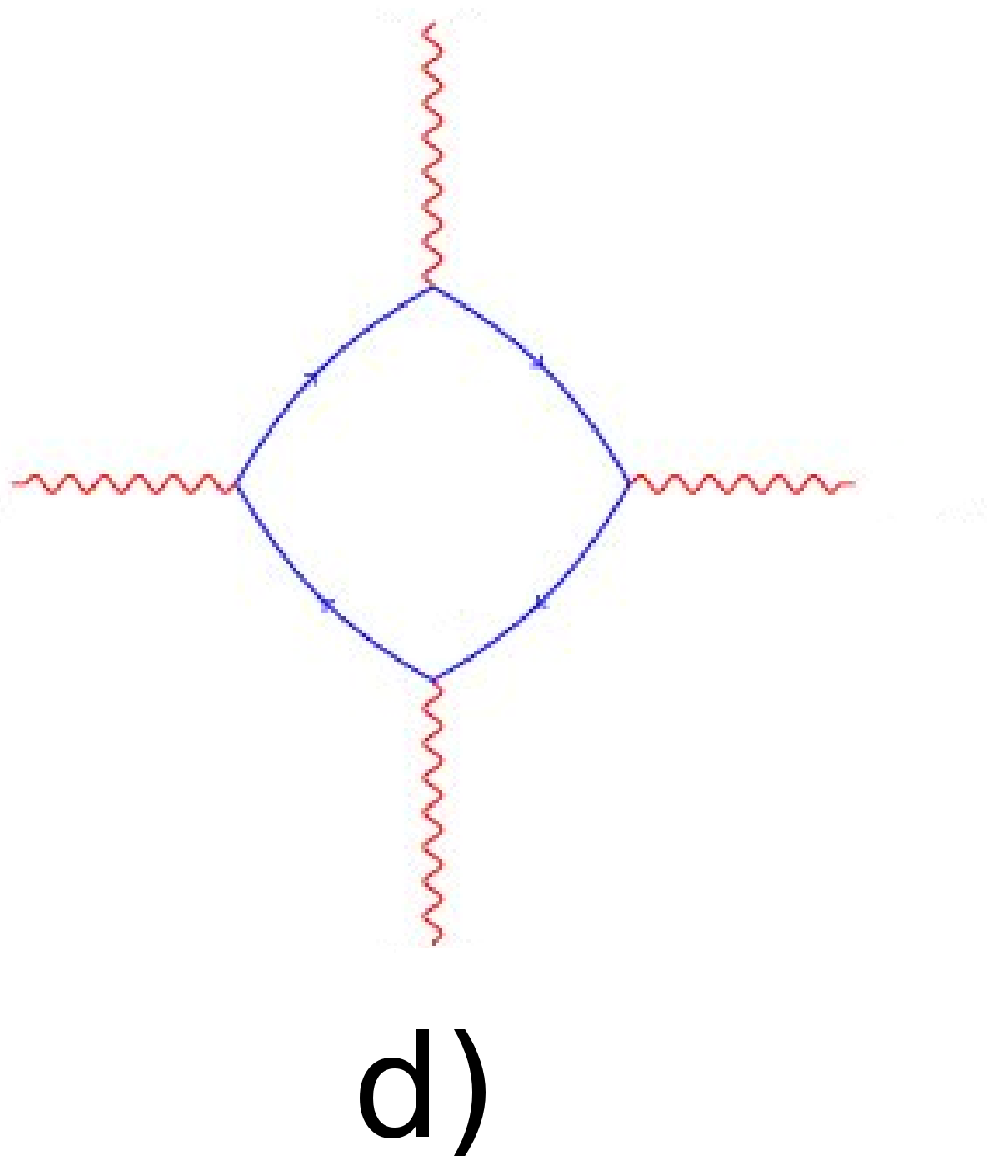}
\caption{$Z \gamma\gamma\gamma\to 0$ process diagrams.\label{FIG2}}
\end{center}
\end{figure}

Block of bosonic diagrams consists of three box diagrams, Fig.\ref{FIG2} (a), six
triangular graphs --- pinches, Fig.\ref{FIG2} (b), and three diagrams of the ``fish'' type --- 
self energies, Fig.\ref{FIG2} (c).

Block of fermionic diagrams consists of only three box diagrams, (Fig.\ref{FIG2} (d).
Each diagram is characterized by different order of 4-momenta of the incoming particle --- $p_4$ 
for $Z$ boson and $p_1, p_2, p_3$ for photons, respectively.

 In the computation of the diagrams --- the application of Feynman rules, the Passarino--Veltman (PV)
reduction of one-loop integrals \cite{11}, scalarizing and separation of the poles --- the expressions of 
the amplitude can be represented in the form of the sum of products Lorentz structures and relevant 
scalar form factors \cite{12}.

In terms of Lorentz structures we write the expression for the covariant amplitude ${Z\gamma\gamma\gamma\to 0}$
\begin{eqnarray}
{\cal A}_{Z\gamma\gamma\gamma\to 0}
  =\sum\limits_{i=1}^{14}\Big[{\cal F}_{i}^{\rm b}\left(s\,,t\,,u\right) 
+ {\cal F}_{i}^{\rm f}\left(s\,,t\,,u\right)\Big] T_{i}^{\alpha\beta\mu\nu}.
\end{eqnarray}
The four rank tensor, with the exclusion by the conservation low of momentum $p_4$ and imposing 
the conditions of physical transversality and zero mass of the photons ($p_{1\alpha}=p_{2\beta}=p_{3\nu}=0$ and
$p^2_{1}=p^2_{2}=p^{2}_{3}=0$), looks like:
\begin{eqnarray}
T^{\alpha\beta\mu\nu}_{1} &=& 
                  \delta_{\alpha\mu}p_{1\beta}p_{2\nu} 
    -             \delta_{\alpha\beta}p_{1\mu}p_{2\nu} 
    -             \delta_{\beta\mu}p_{2\alpha}p_{2\nu} 
\nonumber \\ &&  
    +             \delta_{\beta\nu}p_{1\mu}p_{2\alpha}
    +             \delta_{\beta\nu}p_{2\alpha}p_{2\mu}
    +             \delta_{\mu\nu}p_{2\alpha}p_{3\beta}  
\nonumber \\ &&  
 +\frac{t}{u}\Bigl(
       \delta_{\alpha\nu}p_{1\beta}p_{1\mu}
    -  \delta_{\alpha\beta}p_{1\nu}p_{2\mu}
    -  \delta_{\alpha\mu}p_{1\beta}p_{1\nu}  
\nonumber \\ &&  \hspace*{1cm}
    +  \delta_{\alpha\nu}p_{1\beta}p_{2\mu} 
    +  \delta_{\beta\mu}p_{1\nu}p_{2\alpha}
    +  \delta_{\mu\nu}p_{1\beta}p_{3\alpha}\Bigr)
\nonumber \\ &&  
+ \frac{s}{u}\left(
       \delta_{\alpha\mu}p_{1\nu}p_{3\beta} 
    -  \delta_{\alpha\nu}p_{1\mu}p_{3\beta} 
    +  \delta_{\beta\mu}p_{2\nu}p_{3\alpha} 
    -  \delta_{\beta\nu}p_{2\mu}p_{3\alpha} 
    -  \delta_{\mu\nu}p_{3\alpha}p_{3\beta}\right) 
\nonumber \\ &&  
- \frac{1}{2}\left(  \frac{st}{u}\delta_{\alpha\nu}  \delta_{\beta\mu} 
                               + t\delta_{\alpha\beta}\delta_{\mu\nu}    
                               + s\delta_{\alpha\mu}  \delta_{\beta\nu}\right),
\nonumber \\
T^{\alpha\beta\mu\nu}_{2} &=& 
                 p_{1\beta}p_{1\mu}p_{1\nu}p_{2\alpha} 
            +    p_{1\beta}p_{1\nu}p_{2\alpha}p_{2\mu} 
\nonumber \\ &&
- \frac{1}{2}\left[
                   s\left(\delta_{\alpha\beta}p_{1\mu}p_{1\nu}+\delta_{\alpha\beta}p_{1\nu}p_{2\mu}\right)   
                 - u\delta_{\mu\nu}p_{1\beta}p_{2\alpha}\right] 
- \frac{1}{4}su\delta_{\alpha\beta}\delta_{\mu\nu},
\nonumber\\
 T^{\alpha\beta\mu\nu}_{3} &=&
                     p_{1\beta}p_{1\mu}p_{2\alpha}p_{2\nu} 
      -  \frac{t}{u} p_{1\beta}p_{1\mu}p_{1\nu}p_{2\alpha}
 \nonumber \\ && 
+\frac{1}{2}s\left(\frac{t}{u} \delta_{\alpha\beta}p_{1\mu}p_{1\nu}-\delta_{\alpha\beta}p_{1\mu}p_{2\nu}\right),
\nonumber \\
 T^{\alpha\beta\mu\nu}_{4} &=&
     \frac{t}{u} p_{1\beta}p_{1\mu}p_{1\nu}p_{2\alpha} 
               + p_{1\beta}p_{2\alpha}p_{2\mu}p_{2\nu} 
 \nonumber \\ && 
  - \frac{1}{2}
\left(\frac{st}{u}\delta_{\alpha\beta}p_{1\mu}p_{1\nu} 
               + s\delta_{\alpha\beta}p_{2\mu}p_{2\nu} 
               - t\delta_{\mu\nu}p_{1\beta}p_{2\alpha}\right) 
- \frac{1}{4}st   \delta_{\alpha\beta}\delta_{\mu\nu},
\nonumber\\
 T^{\alpha\beta\mu\nu}_{5} &=&
                p_{1\mu}p_{1\nu}p_{2\alpha}p_{3\beta}
 \nonumber \\ && 
+ \frac{1}{2}
 \bigl[ u \left(  \delta_{\alpha\beta}p_{1\mu}p_{2\nu}
                - \delta_{\alpha\mu}p_{1\beta}p_{2\nu}
                - \delta_{\alpha\mu}p_{1\beta}p_{2\nu}
                - \delta_{\beta\nu}p_{1\mu}p_{2\alpha}\right)
 \nonumber \\ && 
      + t \left(\delta_{\alpha\mu}p_{1\beta}p_{1\nu} - \delta_{\alpha\beta}p_{1\mu}p_{1\nu}\right)
      - s \delta_{\alpha\mu}p_{1\nu}p_{3\beta}   
   \bigr]
+ \frac{1}{4}su\delta_{\alpha\mu}\delta_{\beta\nu},
\nonumber\\
 T^{\alpha\beta\mu\nu}_{6} &=&
              p_{1\nu}p_{2\alpha}p_{2\mu}p_{3\beta}  
 - \frac{1}{2} \bigl[
s\left(
 \delta_{\beta\mu}p_{2\nu}p_{3\alpha}
-\delta_{\beta\nu}p_{2\mu}p_{3\alpha}
+\delta_{\alpha\nu}p_{2\mu}p_{3\beta} \right)        
 \nonumber \\ &&    
+ t\delta_{\beta\mu}p_{1\nu}p_{2\alpha} 
- u\left(\delta_{\beta\mu}p_{2\alpha}p_{2\nu}-\delta_{\beta\nu}p_{2\alpha}p_{2\mu}\right)   
               \bigr]
+ \frac{1}{4} st\delta_{\alpha\nu}\delta_{\beta\mu}, 
\nonumber \\
 T^{\alpha\beta\mu\nu}_{7} &=&
 p_{1\mu}p_{2\alpha}p_{2\nu}p_{3\beta}     
- \frac{1}{2}\left(s\delta_{\alpha\mu}p_{2\nu}p_{3\beta} + t\delta_{\beta\nu}p_{1\mu}p_{2\alpha}\right) 
+ \frac{1}{4}st\delta_{\alpha\mu}\delta_{\beta\nu},
\nonumber \\
 T^{\alpha\beta\mu\nu}_{8} &=&
 p_{2\alpha}p_{2\mu}p_{2\nu}p_{3\beta} 
- \frac{s}{u}p_{2\mu}p_{2\nu}p_{3\alpha}p_{3\beta} 
- \frac{1}{2}t\left(\delta_{\beta\nu}p_{2\alpha}p_{2\mu}-\frac{s}{u}\delta_{\beta\nu}p_{2\mu}p_{3\alpha}\right),
\nonumber \\
 T^{\alpha\beta\mu\nu}_{9} &=&
             p_{1\beta}p_{1\mu}p_{1\nu}p_{3\alpha} 
- \frac{s}{t}p_{1\mu}p_{1\nu}p_{3\alpha}p_{3\beta}  
- \frac{1}{2}u\left(\delta_{\alpha\nu}p_{1\beta}p_{1\mu}-\frac{s}{t}\delta_{\alpha\nu}p_{1\mu}p_{3\beta}\right),
\nonumber \\
 T^{\alpha\beta\mu\nu}_{10} &=&
 p_{1\beta}p_{1\nu}p_{2\mu}p_{3\alpha} 
-   \frac{1}{2}\left(u\delta_{\alpha\nu}p_{1\beta}p_{2\mu}-s\delta_{\beta\mu}p_{1\nu}p_{3\alpha}\right)
+ su\frac{1}{4}\delta_{\alpha\nu}\delta_{\beta\mu},
\nonumber\\
 T^{\alpha\beta\mu\nu}_{11} &=&
 p_{1\beta}p_{1\mu}p_{2\nu}p_{3\alpha}
+  \frac{1}{2}\bigl[
- s\left(\delta_{\alpha\nu}p_{1\mu}p_{3\beta}-\delta_{\alpha\mu}p_{1\nu}p_{3\beta}-\delta_{\beta\nu}p_{1\mu}p_{3\alpha}\right)
\nonumber \\ &&
+ t\left(\delta_{\alpha\mu}p_{1\beta}p_{1\nu}-\delta_{\alpha\nu}p_{1\beta}p_{1\mu}\right)   
- u \delta_{\alpha\mu}p_{1\beta}p_{2\nu}\bigr]
+ \frac{1}{4}su\delta_{\alpha\mu}\delta_{\beta\nu},
\nonumber \\
 T^{\alpha\beta\mu\nu}_{12} &=&
    p_{1\beta}p_{2\mu}p_{2\nu}p_{3\alpha} 
- \frac{1}{2}s\delta_{\beta\mu}p_{2\nu}p_{3\alpha}
\nonumber \\ &&
+ \frac{1}{2}t\left(\delta_{\alpha\beta}p_{1\nu}p_{2\mu}-\delta_{\alpha\nu}p_{1\beta}p_{2\mu}-\delta_{\beta\mu}p_{1\nu}p_{2\alpha}\right)     
\nonumber \\ &&
+ \frac{1}{2}u\left(\delta_{\beta\mu}p_{2\alpha}p_{2\nu}-\delta_{\alpha\beta}p_{2\mu}p_{2\nu}\right)   
+ \frac{1}{4}st\delta_{\alpha\nu}\delta_{\beta\mu}, 
\nonumber\\
 T^{\alpha\beta\mu\nu}_{13} &=&
              p_{1\nu}p_{2\mu}p_{3\alpha}p_{3\beta} 
- \frac{1}{2}\left( t\delta_{\beta\mu}p_{1\nu}p_{3\alpha}+u\delta_{\alpha\nu}p_{2\mu}p_{3\beta}\right)
+ \frac{1}{4}tu\delta_{\alpha\nu}\delta_{\beta\mu},
\nonumber \\
 T^{\alpha\beta\mu\nu}_{14} &=&
               p_{1\mu}p_{2\nu}p_{3\alpha}p_{3\beta} 
- \frac{1}{2}  \left(t\delta_{\beta\nu }p_{1\mu}p_{3\alpha}+u\delta_{\alpha\mu}p_{2\nu}p_{3\beta}\right)
+ \frac{1}{4} tu\delta_{\alpha\mu}\delta_{\beta\nu}.
\nonumber
\end{eqnarray}

\begin{figure}[ht!]
\begin{center}
\includegraphics[width=4.5cm,angle=0]{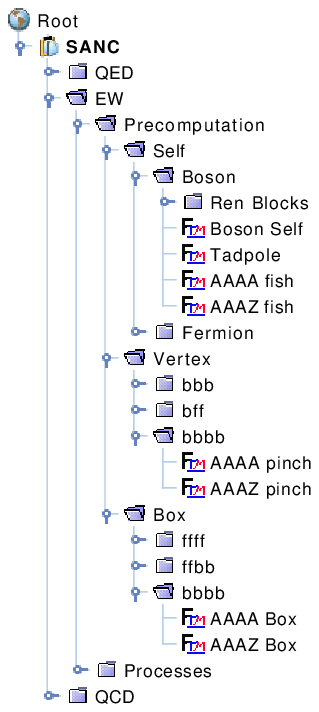}
\includegraphics[width=4.5cm,angle=0]{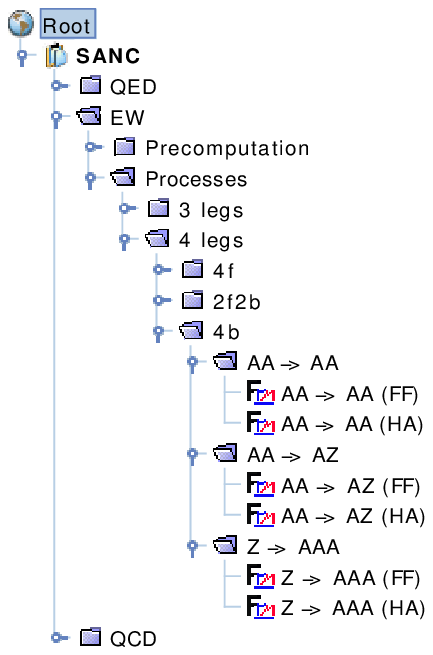}
\caption{$Z\gamma\gamma\gamma\to 0$ {\tt SANC} processes tree.\label{FIG3}}
\end{center}
\end{figure}

The precomputation files {\tt AAAZ Box, AAAZ pinch, AAAZ fish} (see the {\tt SANC} process tree in Fig.\ref{FIG3} from {\tt SANC} client \cite{3}) contains the sequence of procedures for calculation of the covariant amplitude. 

Form factors ${\cal F}_{i}$ are the scalar coefficients in front of basis structures of the covariant amplitude. 
They are presented as some combinations of scalar PV functions $A_0$, $B_0$, $C_0$, $D_0$~\cite{11}, 
and depend on invariants $s\,,t\,,u\,,$ and also on fermion and boson masses. They do not contain 
ultraviolet poles.
The derived one-loop scalar form factors can be used for any cross channel after an appropriate
permutation of their arguments $s,t,u$.

Explicit expressions for the boson and fermion parts of the forms factors are not shown in this article  
because they are very cumbersome.
A complete answer for ${\cal F}_{i}$ one can be found in the package which is downloadable from the 
homepages of the computer system {\tt SANC}.
Note that the expression for the amplitude of boson diagrams are similar to fermion 
one except for the explicit representation of form factors.

To the Lorentz structure of the expression, the axial interaction of $Z$-boson with fermions 
$g^Z_{Af} \epsilon_{\alpha \beta \nu \mu}$ does not give the contribution due to charge symmetry. One can be
shown that it cancels in a full set of diagrams.

For the analytical proof of this fact a special sequence of the Shouten identities was used:
\begin{eqnarray}
\epsilon_{\mu_1? \mu_2? \mu_3? \mu_4?}\delta_{\mu_5?\mu_6?} &=&
 \epsilon_{\mu_5 \mu_2 \mu_3 \mu_4} \delta_{\mu_1\mu_6}
+\epsilon_{\mu_1 \mu_5 \mu_3 \mu_4} \delta_{\mu_2\mu_6}
\\ \nonumber &+& 
   \epsilon_{\mu_1 \mu_2 \mu_5 \mu_4} \delta_{\mu_3\mu_6}
  +\epsilon_{\mu_1 \mu_2 \mu_3 \mu_5} \delta_{\mu_4\mu_6},
\label{basicShooten}
\end{eqnarray}
where $\mu_i?$ denotes {\it any} index.

Namely, we applied, step-by-step, the basic identity~\ref{basicShooten} contracted with certain  
number of 4-momenta.
It was found only 5 families of the lhs of the Shouten identities for the substitutions. 
\begin{itemize}
\item{}
Family $\epsilon_{p_1 p_2 p_3 \alpha?} \delta_{\mu_5? \mu_6?}$
has~~4~~members: \\
$\epsilon_{p_1 p_2 p_3 \alpha} \delta_{\mu_5?,\mu_6?}$, 
~$\epsilon_{p_1 p_2 p_3 \beta} \delta_{\alpha \mu_5?}$, 
~$\epsilon_{p_1 p_2 p_3 \beta} \delta_{\nu \mu_6?}$, 
~$\epsilon_{p_1 p_2 p_3 \mu} \delta_{\nu \mu_6?}$. 

\item{}
Family $\epsilon_{p_1?,p_2?,\alpha?,\beta?} \delta_{\mu?,\nu?}$
has~~3~~members: \\ 
~$\epsilon_{p_1? p_2? \alpha \beta} \delta_{\mu\nu}$, 
~$\epsilon_{p_1? p_2? \alpha \nu} \delta_{\mu\beta}$,  
~$\epsilon_{p_1? p_2? \mu \nu}    \delta_{\alpha\beta}$.

\item{}
Family  $\epsilon_{p_i,p_j,\alpha?,\nu?} p_{k\mu?}$ has 11~~members with 3~~sets each\\
(first set for $i,j=1,2$, second set for $i,j=1,3$ and the last one for $i,j=3,2$):\\
 $\epsilon_{p_i p_j \alpha? \nu} p_{3\beta}$,
 $\epsilon_{p_ ,p_j \alpha? \mu} p_{3\beta}$, 
 $\epsilon_{p_i p_j \beta? \nu}  p_{3\alpha}$,\\ 
 $\epsilon_{p_i p_j \alpha \nu}  p_{1\mu} p_{3\beta}$, 
 $\epsilon_{p_i p_j \alpha \mu}  p_{1\nu} p_{3\beta}$, 
 $\epsilon_{p_i p_j \beta \mu}   p_{2\nu} p_{3\alpha}$,
 $\epsilon_{p_i p_j \beta \nu}   p_{2\mu} p_{3\alpha}$, \\
 $\epsilon_{p_i p_j \beta \nu?}  p_{2\alpha}$, 
 $\epsilon_{p_i p_j \alpha \nu?} p_{1\beta}$,
 $\epsilon_{p_i p_j \mu \nu}    p_{1\beta}$,
 $\epsilon_{p_i p_j \beta \nu?}  p_{2\alpha}$,

\item{
Family~~name $\epsilon_{p_i,?,?,?} p_{k\mu?}$ has 3~~members with 3 sets (for $i=1,2,3$):\\
$\epsilon_{p_i \beta  \mu \nu}  p_{3 \alpha}$,
$\epsilon_{p_i \beta  \mu \nu}  p_{2 \alpha}$,
$\epsilon_{p_i \alpha \mu \nu } p_{1 \beta} $.}
\end{itemize}

The procedure {\tt  Projections()} was call in between. 

\begin{verbatim}
#procedure Projections()
id p1(al)=0;
id p2(be)=0;
id p3(nu)=0;
id p3(mu)=-p1(mu)-p2(mu);

repeat id p1.p1=0;
repeat id p2.p2=0;
repeat id p3.p3=0;
repeat id p1.p2=-s/2;
repeat id p1.p3=-u/2;
repeat id p2.p3=-t/2;
#endprocedure
\end{verbatim}

\section{\label{SEC3} Processes level,~ helicity amplitude.}
When we implement any processes, we create the building block for annihilation to the vacuum and then
use  these building blocks several times replacing incoming momenta $p$'s 
by corresponding kinematic momenta with the right signs. 

The covariant amplitude for the channel 
$\gamma\gamma\rightarrow \gamma Z$ can be obtained from annihilation to the vacuum with
the following permutation of the 4-momenta:
\begin{eqnarray}
&& p_1 \to \phantom{-} p_1, \\ \nonumber
&& p_2 \to \phantom{-} p_2, \\ \nonumber
&& p_3 \to          -  p_3, \\ \nonumber
&& p_4 \to          -  p_4     \nonumber 
\end{eqnarray}
and for the decay $Z \to \gamma\gamma\gamma$ are:
\begin{eqnarray}
&& p_1 \to          -  p_1, \\ \nonumber
&& p_2 \to          -  p_2, \\ \nonumber
&& p_3 \to          -  p_3, \\ \nonumber
&& p_4 \to \phantom{-} p_4.    \nonumber
\end{eqnarray}

Furthermore we calculate ${\cal F}_{i}$ by module {\tt AA $\to$ AZ (FF)}, {\tt Z $\to$ AAA (FF)}, 
then helicity amplitudes by the module {\tt AA $\to$ AZ (HA)}, {\tt Z $\to$ AAA (HA)} and finally --- the analytic expression for differential and total process cross section in sanc\_4b\_v1.00 package.

For more compact presentation of the results and more effective numerical implementation in 
the system {\tt SANC} it was applied the method of helicity amplitudes.
Lorentz structure of the classical expression is contracted with polarization vectors, 
and one gets the orthogonal set of scalar variables, expressed in terms of forms factors --- 
helicity amplitudes \cite{13}.

The expressions for helicity amplitude with substitution of boson and fermion forms factors for both 
channels were calculated. It was received the full agreement as compared with the calculated in the literature \cite{4,5,6,7,8} (though some trivial misprints had to be corrected).

The cross section of the reaction $\gamma\gamma\to\gamma Z$ is calculated by the formulae:
\begin{eqnarray}
d\sigma_{\gamma\gamma\to\gamma Z}=
\frac{1}{4\sqrt{(p_1p_2)^2}}\left|{\cal{A}}_{\gamma\gamma\to\gamma Z}\right|^2 d\Phi^{(2)},
\nonumber
\end{eqnarray}
where ${\cal{A}}_{\gamma\gamma\to\gamma Z}$ is the covariant amplitude of the process and $d\Phi^{(2)}$ is the 
two-body phase space:
\begin{eqnarray}
d\Phi^{(2)}=(2\pi)^4\delta\left(p_1+p_2-p_3-p_4\right) 
\frac{d^4 p_3 \delta\left( p_3^2 \right)}{(2\pi)^3}\frac{d^4p_4\delta\left( p_4^2 \right)}{(2\pi)^3}.
\nonumber
\end{eqnarray}

For the differential cross section one gets: 
\begin{eqnarray}
d\sigma_{\gamma\gamma\to\gamma Z}=
\frac{1}{32\pi s}\left(1-\frac{M_Z^2}{s}\right)\left|{\cal{A}}_{\gamma\gamma\to\gamma Z}\right|^2 d \cos\theta,
\nonumber
\end{eqnarray}
where $\theta$ is the scattering angle of the Z boson in the center of mass system.

The decay width of the $Z$ boson is calculated by the formula:
\begin{eqnarray}
\Gamma_Z = \frac{1}{3! 384 \pi^3 M_Z^3} 
\int \left|{\cal{A}}_{Z\to\gamma\gamma\gamma}\right|^2 ds\, dt\, du \times \delta(M_Z^2-s-t-u).
\nonumber
\end{eqnarray}

\section{\label{SEC4} Technical description of the package}
The process ~(\ref{Reaction}) and the decay ~(\ref{Decay}) and also previously implemented process
$\gamma\gamma\to\gamma\gamma$ are included into the package {\rm sanc\_4b\_v1.00} (Fig.\ref{FIG3}),
which can be downloaded from the project site \cite{3}.

Our main purpose is to introduce the package {\rm sanc\_4b\_v1.00} into generic integrator SANC,  
based on an algorithm VEGAS \cite{14}.

Here we present the technical description of this package --- main flags and the options:
\begin{enumerate}
\item {\rm bbbb\_main.F} --- the main file,
\item {\rm bbbb\_ha\_11\_11.F} --- the HA of $\gamma\gamma\to\gamma\gamma$ process from {\tt SANC} system,
\item {\rm bbbb\_ha\_11\_12.F} --- the HA of $\gamma\gamma\to\gamma Z$ process from {\tt SANC} system,
\item {\rm bbbb\_ha\_2\_111.F} --- the HA of $Z\to\gamma\gamma\gamma$ process from {\tt SANC} system,
\item {\rm *.f} --- the library of special functions and algorithms,
\item {\rm *\_input.h} --- the set of various setups of input parameters,
\item {\rm README, INSTALL} and other instructions files.
\end{enumerate}

In {\rm README} and {\rm INSTALL} files one can find instructions how to use the package. 
The main options one can change in {\rm bbbb\_main.F}.

\vspace{2mm}

{\tt\bf pid(I)} --- choice of the process:
\begin{itemize}
\item{I = AA2AA, \qquad $\gamma\gamma\to\gamma\gamma$ process} 
\item{I = AA2AZ, \qquad $\gamma\gamma\to\gamma Z$ process} 
\item{I = Z2AAA, \qquad $Z \to\gamma\gamma\gamma$ decay}
\end{itemize}
\vspace{2mm}

{\tt\bf ipm(I)} --- choice of incoming photons helicities sum in cross section:
\begin{itemize}
\item{I = SS, \qquad  total helicities sum} 
\item{I = $++$, \qquad $"++"$ helicities sum} 
\item{I = $+-$, \qquad $"+-"$ helicities sum}
\end{itemize}
\vspace{2mm}

{\tt\bf itl(I)} --- choice of helicities for Z-boson:
\begin{itemize}
\item{I = T, \qquad Z-transverse helicities} 
\item{I = L, \qquad Z-longitudinal helicities} 
\item{I = S, \qquad Z-helicities sum}
\end{itemize}
\vspace{2mm}

{\tt\bf iqed(I)} --- choice of calculations for QED (fermionic) corrections:
\begin{itemize}
\item{I = 0, \qquad without QED corrections} 
\item{I = 1, \qquad with QED corrections}
\end{itemize}
\vspace{2mm}

{\tt\bf iew(I)} --- choice of calculations for EW (bosonic) corrections:
\begin{itemize}
\item{I = 0, \qquad without EW corrections} 
\item{I = 1, \qquad with EW corrections}
\end{itemize}
To get full EW answer with the interference one should set {\tt iqed = 1, \tt iew = 1.}

\vspace{2mm}

{\tt\bf gfscheme(I)} --- choice of the EW scheme:
\begin{itemize}
\item{I = 0, \qquad $\alpha_0$ calculation scheme} 
\item{I = 1, \qquad $G_F$ scheme} 
\item{I = 2, \qquad ${G'}_F$ scheme, 
\\ \hspace*{2cm} when $\alpha_0$ is replaced by $\alpha_{G_F}=\sqrt{2}G_F M^2_W \left(1-M^2_W / M^2_Z \right) / \pi.$}
\end{itemize}             
\vspace{2mm}

{\tt\bf isetup(I)} --- choice of the setup:
\begin{itemize}
\item{I = 0, \qquad Standard {\tt SANC} input [PDG 2006]}
\item{I = 1, \qquad Les Houches Workshop (2005)}
\item{I = 2, \qquad Tevatron-for-LHC Workshop (2006)}
\item{I = 3, \qquad Custom setup}
\end{itemize}

\section{\label{SEC5} Numerical results}
The calculations were made for the cross-section of the reaction $\gamma\gamma\to\gamma Z$ 
and decay width of the $Z\to\gamma\gamma\gamma$ with the following values of parameters:
\begin{eqnarray}
&& \alpha = 1/128; \nonumber\\
&& \pi/6 < \theta < 5\pi/6; \nonumber\\
&& M_W = 80.22~{\rm GeV;} \nonumber\\
&& M_Z = 91.173~{\rm GeV;} \nonumber\\
&& m_e = 0.1~{\rm GeV,}~~~m_\mu= 0.1~{\rm GeV,}~~~m_\tau = 0.1~{\rm GeV;} \nonumber\\
&& m_u = 0.1~{\rm GeV,}~~~m_c = 0.1~{\rm GeV,}~~~m_t = 120.0~{\rm GeV;} \nonumber\\
&& m_d = 0.1~{\rm GeV,}~~~m_s = 0.1~{\rm GeV,}~~~m_b = 5.0~{\rm GeV.}
\end{eqnarray}

\begin{figure}[!ht]
\begin{center}
\includegraphics[width=6cm,angle=270]{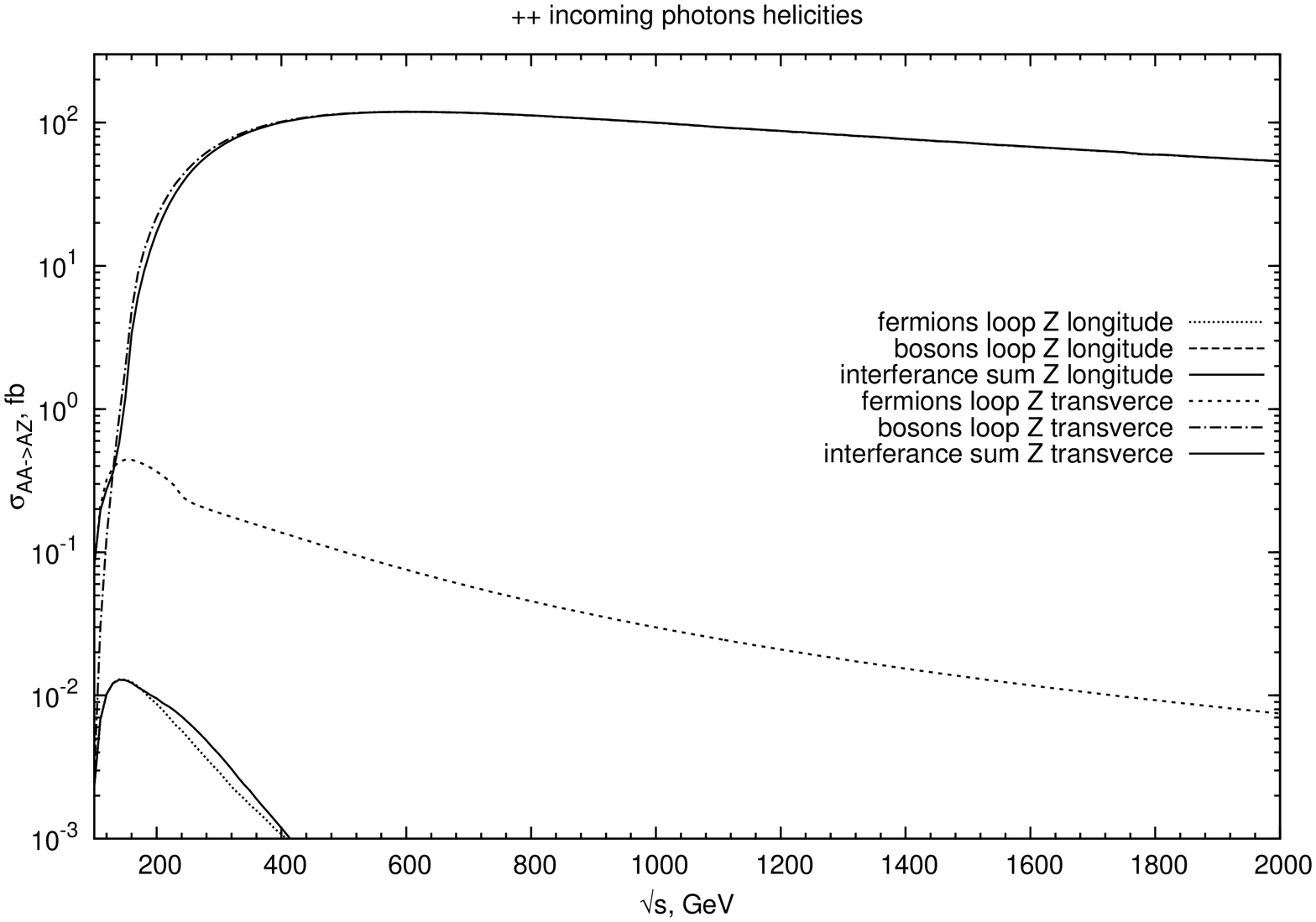}
\caption{$\gamma\gamma\to\gamma Z$ {\tt SANC} cross section ($"++"$) incoming photons helicities.\label{FIG4}}
\end{center}
\end{figure}

\begin{figure}[!ht]
\begin{center}
\includegraphics[width=6cm,angle=270]{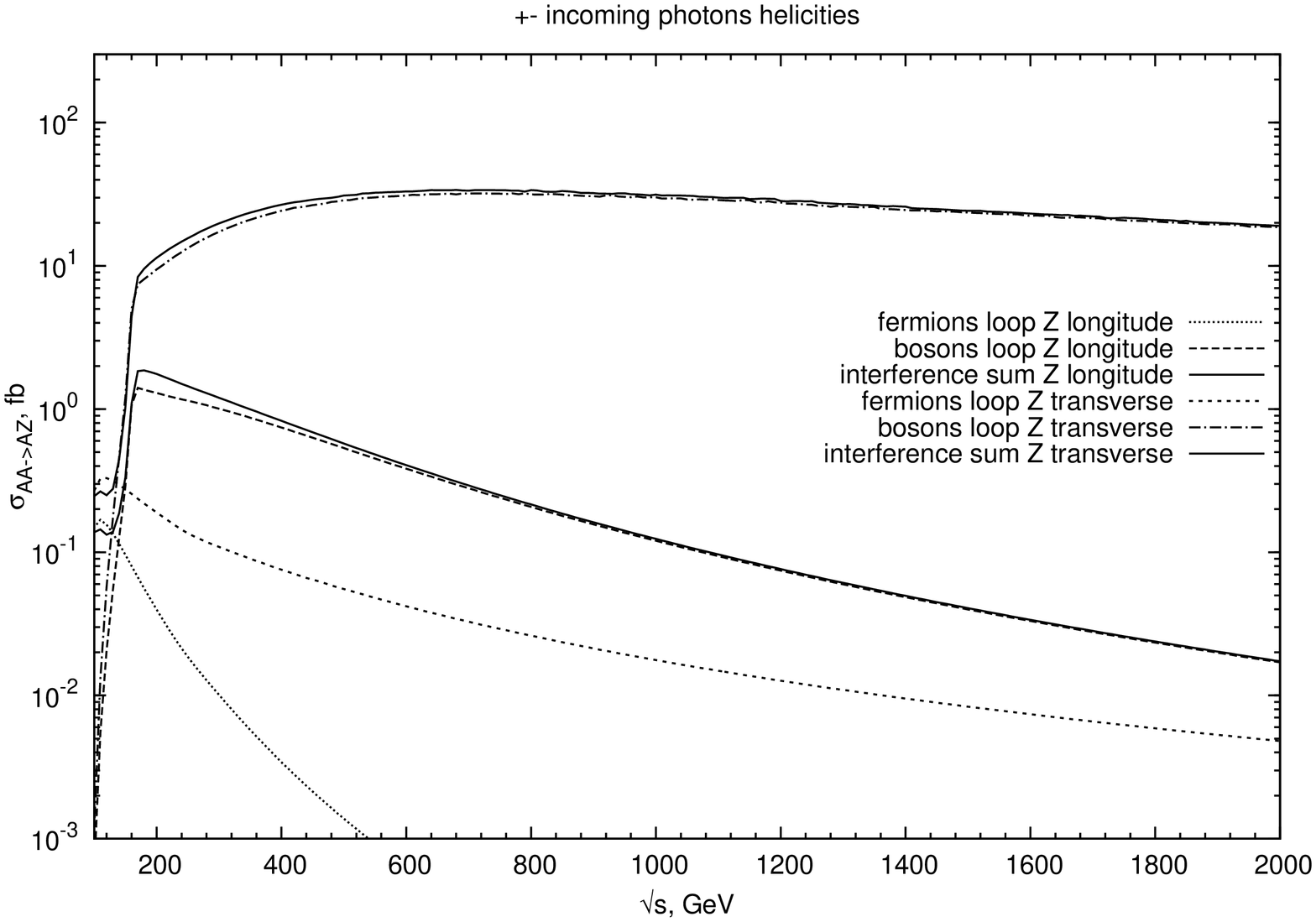}
\caption{$\gamma\gamma\to\gamma Z$ {\tt SANC} cross section ($"+-"$) incoming photons helicities.\label{FIG5}}
\end{center}
\end{figure}

The numerical results for the reaction $\gamma\gamma\to\gamma Z$ were compared to \cite{4,5,6,7,8} componentwise
for boson, fermion contributions and their interference taking into account  the helicities  of $Z$ boson, and
initial photons (for $++$ see Fig.\ref{FIG4}, for $+-$ see Fig.\ref{FIG5}) in the energy 
range from 100 GeV up to 2 TeV.

For all of the contributions it was obtained a good agreement with the results, given in the literature.

\begin{figure}[!ht]
\begin{center}
\includegraphics[width=7cm,angle=270]{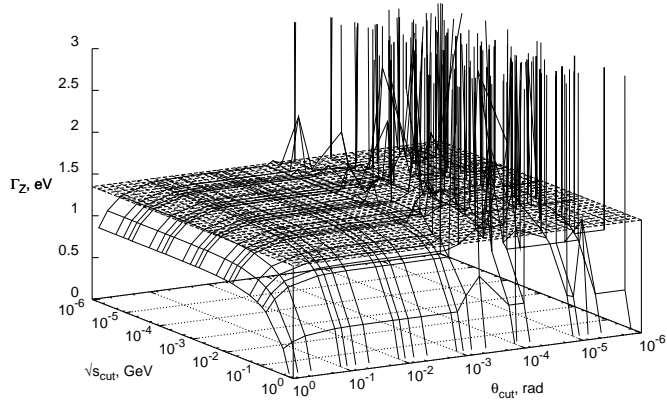}
\caption{$Z \to\gamma\gamma\gamma$ {\tt SANC} decay width plateau of stability.\label{FIG6}}
\end{center}
\end{figure}

When calculating the decay width $\Gamma_Z$, the variation of the cut parameters 
(the angle $\theta_{cut}$ and the energy of photons $\sqrt s_{cut}$) was performed in a large range of values (Fig.\ref{FIG6}). 
It was found a wide plateau of stability, giving a result consistent with those given in the literature \cite{7}.

\section{\label{SEC6} Conclusions}

 The precomputation strategy of the {\tt SANC} system \cite{1} and the place of the process
(\ref{Reaction}) and (\ref{Decay}) on the {\tt SANC} process tree were presented.

The implementation of analytical results and the concept of modules were described.

Its numerical results of the sanc\_4b\_v1.00 package were compared with those 
existing in the literature. The package is available for download at web page \cite{3}. 

The authors are grateful to ~S.~Bondarenko, G.~Nanava, V.~Kolesnikov, and A.~Sapronov for useful discussions of 
numerical and analytical calculations.

\newpage

%
\end{document}